\documentclass[a4paper,11pt]{article}

\usepackage{amsmath,amssymb,amscd,bbm,amsfonts}
\usepackage{graphicx}
\usepackage{multirow}
\usepackage{cite}
\makeatletter

 \@addtoreset{equation}{section}
\makeatother

\newcounter{Enumerate}

\DeclareFontFamily{U}{rsf}{}
\DeclareFontShape{U}{rsf}{m}{n}{
  <5> <6> rsfs5 <7> <8> <9> rsfs7 <10-> rsfs10}{}
\DeclareMathAlphabet\Scr{U}{rsf}{m}{n}

\usepackage[mathscr]{eucal}

\newcommand{\del}{\partial}
\newcommand{\half}{\frac{1}{2}}

\newcommand{\LS}{\ \ \ \ \ \ \ \ \ \ }
\newcommand{\ls}{\ \ \ \ \ }
\newcommand{\wt}{\widetilde}
\newcommand{\wh}{\widehat}

\newcommand{\ve}{\varepsilon}
\newcommand{\ol}{\overline}

\newcommand{\bsubeq}{\begin{subequations}}
\newcommand{\esubeq}{\end{subequations}}

\renewcommand{\d}{{\rm d}}
\newcommand{\nn}{\nonumber}
\newcommand{\e}{{\rm e}}
\newcommand{\slb}{\scalebox}
\newcommand{\VBH}{I_1}
\newcommand{\rH}{r_{\text{H}}}
\newcommand{\rA}{r_{\text{A}}}
\newcommand{\VE}{V_{\text{eff}}}
\newcommand{\GV}{G_{\text{V}}}
\newcommand{\T}{{\rm T}}

\makeatletter
\def\bcline#1{\@bcline#1\@nil}
\def\@bcline#1-#2\@nil{%
  \omit
  \@multicnt#1%
  \advance\@multispan\m@ne
  \ifnum\@multicnt=\@ne\@firstofone{&\omit}\fi
  \@multicnt#2%
  \advance\@multicnt-#1%
  \advance\@multispan\@ne
  \cleaders\hbox{$\m@th \mbox{\rule{.1em}{.035em}}\mkern2mu$}\hfill
  \cr
  \noalign{\vskip-\arrayrulewidth}}
\def\multispan{\omit\@multispan}
\def\@multispan#1{%
  \@multicnt#1\relax
  \loop\ifnum\@multicnt>\@ne \sp@n\repeat}
\def\sp@n{\span\omit\advance\@multicnt\m@ne}
\makeatother

\begin{document}
\allowdisplaybreaks{

\thispagestyle{empty}


\begin{flushright}
KEK-TH-1367 \\
\end{flushright}

\vspace{25mm}

\begin{center}
\slb{1.75}{Non-supersymmetric Extremal RN-AdS Black Holes}

\vspace{3mm}

\slb{1.75}{in ${\cal N}=2$ Gauged Supergravity}

\vspace{15mm}

\slb{1.2}{Tetsuji {\sc Kimura}}

\vspace{2mm}

{\sl
KEK Theory Center,
Institute of Particle and Nuclear Studies, \\
High Energy Accelerator Research Organization \\
Tsukuba, Ibaraki 305-0801, Japan}

\vspace{1mm}

\slb{0.9}{\tt tetsuji@post.kek.jp}

\end{center}

\vspace{18mm}


\begin{abstract}
We investigate extremal Reissner-Nordstr\"{o}m-AdS black holes in four-dimensional ${\cal N}=2$ abelian gauged supergravity. 
We find a new attractor equation which is not reduced to the one in the asymptotically flat spacetime. 
We also argue a formula which is available even in the presence of the scalar potential.
We apply them to the T$^3$-model and the STU-model in generic black hole charge distributions.
In addition, focusing on the so-called T$^3$-model with a single neutral 
vector multiplet, 
we obtain non-supersymmetric extremal Reissner-Nordstr\"{o}m-AdS black hole solutions with regular event horizons in the D0-D4 and the D2-D6 black hole charge configurations. 
The negative cosmological constant emerges even without the Fayet-Iliopoulos parameters. 
\end{abstract}

\newpage

\section{Introduction}

Quantum field theory with eight supersymmetry charges has been intensively studied both in perturbative and non-perturbative viewpoints.
Moduli spaces in a global (or local) supersymmetric theory with eight supercharges are governed by highly established geometries such as special K\"{a}hler geometry and as hyper-K\"{a}hler (or quaternionic) geometry.
By virtue of mathematical feature of these geometries, 
four-dimensional gauge theory \cite{Seiberg:1994rs, Seiberg:1994aj} and 
supergravity \cite{Andrianopoli:1996cm} have been explored exhaustively.
They have also been further developed in string theory
in a vast literature on spacetime compactification scenarios.

In string theory compactification on a certain geometry such as a Calabi-Yau variety, however, a four-dimensional effective theory does not involve a (non-abelian) gauge symmetry which must emerge in a realistic model. 
In order for the gauge symmetry to be naturally generated, non-trivial fluxes or twisting on the internal space is inevitable. 
Mutually supported with string theory, 
gauged supergravities in diverse dimensions with the extended number of supercharges have been investigated (see, for instance, an instructive lecture \cite{Samtleben:2008pe} and references therein).
The gauge symmetry yields a non-trivial potential term whose expectation value becomes the cosmological constant.
This indicates that  gauged supergravity gives rise to a resolution of the cosmological problem.

Extended supergravity provides another physical perspective: a search of (non-)extremal black hole solutions.
In the absence of the cosmological constant, a remarkable feature called the attractor mechanism can be discussed.
In the case of extremal charged black holes, referred to as the extremal Reissner-Nordstr\"{o}m (RN) black holes, 
the Bekenstein-Hawking entropy is realized in terms of the black hole charges 
without any pieces of physical information at infinity.
A typical situation is a static, spherically symmetric, charged black hole in the asymptotically flat spacetime. 
In this case both BPS and non-BPS solutions have been studied and classified (see, for instance, \cite{Ferrara:1995ih, Ferrara:1997tw, Goldstein:2005hq, Bellucci:2006xz, Kallosh:2006ib, Bellucci:2006zz, Nampuri:2007gv} and references therein). 
Even in non-static configurations, extremal multi-centered black holes
have also been analyzed in BPS \cite{Denef:2007vg} and non-BPS configurations \cite{Gaiotto:2007ag}.

A search of (extremal) black holes in asymptotically anti-de Sitter (AdS) spacetime configurations should also be notable
in the context of AdS/CFT correspondence as well as the black hole physics itself.
The bare negative cosmological constant can be naturally involved if the system possesses neither vector multiplets nor hypermultiplets \cite{Freedman:1976aw}. 
By using this system various supersymmetric black hole solutions are classified \cite{Caldarelli:1998hg}.
Analyses in the presence of vector multiplets and/or hypermultiplets should be explored as in the case of charged black holes in the asymptotically non-flat spacetime.
Notice that, if the spacetime is non-rotating and asymptotically AdS whose event horizon has a certain topology, a supersymmetric solution has a naked singularity in order to satisfy vanishing supersymmetry variations of gravitini. 
The naked singularity emerges in the electric \cite{Romans:1991nq}, magnetic, 
and dyonic black holes \cite{Chamseddine:2000bk} in the four-dimensional asymptotically AdS spacetime. 
Recently some non-supersymmetric extremal RN-AdS black hole solutions were inspected in abelian gauged supergravity in the presence of the Fayet-Iliopoulos (FI) parameters \cite{Bellucci:2008cb}. 
The FI parameter triggers the emergence of the non-vanishing cosmological constant even in the presence of vector multiplets.

It is also interesting to search extremal AdS black holes in the absence of the FI parameters.
In this case the emergence of the non-vanishing cosmological constant would not be guaranteed. 
It is rather difficult to realize such a situation compared to the asymptotically flat case.
In many cases the scalar potential vanishes at the horizon.
In this work we study the existence of the extremal RN-AdS black hole even without the FI parameters. 
We derive a new attractor equation from the equations of motion 
under various severe constraints. 
This is not smoothly connected to the one in the flat case.
Generically a nonlinear function appears in the attractor equation.
In a generic distribution of the black hole charges,
we obtain a description of non-supersymmetric solution of the extremal RN-AdS black hole both in the T$^3$-model and in the STU-model.
We utilize a partial differential equation for the symplectic invariant \cite{Kallosh:2006ib}.
This is irrespective of the scalar potential.
Combining it with the new attractor equation,
we encounter nonlinear solutions.
Next, 
we heuristically analyze some examples in the T$^3$-model.
We introduce only a single 
vector multiplet without any hypermultiplets.
We also restrict the black hole charge distributions to the magnetic configuration, i.e., the D0-D4 system, and to the D2-D6 system.
Even in such a strong restriction we obtain a non-supersymmetric solution which involves the non-trivial black hole entropy and the negative cosmological constant.
Different from the flat case, the entropy is not simply described in terms of the discriminant of the holomorphic central charge.
This solution never smoothly approaches to the non-BPS solution in the flat spacetime. 

This paper is organized as follows:
In section \ref{setup} we review abelian gauged supergravity involving the scalar potential. We exhibit the equations of motion for bosonic fields. 
Some features of the vector moduli space as the special K\"{a}hler geometry are introduced.
In section \ref{main} we investigate the existence of non-supersymmetric solutions of the extremal RN-AdS black hole.
First, in section \ref{metric_ansatz}, we explore a black hole metric ansatz in the asymptotically AdS spacetime. 
We focus on a static, spherically symmetric, charged extremal black hole metric. 
Second, in \ref{effective_potential}, we derive the new attractor equation which is not reduced to the one in the asymptotically flat spacetime.
Third, in section \ref{search}, we classify situations which have to satisfy the new attractor equation.
We find two examples of the extremal RN-AdS black hole in section \ref{examples}.
Section \ref{discussions} is devoted to discussions.
In appendix \ref{non-BPS_RN} we briefly review a non-BPS black hole solution in the asymptotically flat spacetime.


\section{Abelian gauged supergravity with vector multiplets}
\label{setup}

\subsection{Lagrangian and equations of motion}

In the present work we only focus on four-dimensional ${\cal N}=2$ abelian gauged supergravity which contains $n_V$ neutral 
vector multiplets in the absence of hypermultiplets \cite{Andrianopoli:1996cm}. 
Because of gauging, the system has a non-trivial scalar potential which depends on scalar fields of the vector multiplets.
Here we exhibit the bosonic part of the action:
\begin{align}
S \ &= \ 
\int \d^4 x \sqrt{-g} \left\{
\frac{1}{2 \kappa^2} R - g^{\mu \nu} G_{a \ol{b}} \del_{\mu} z^a \del_{\nu} \ol{z}{}^{\ol{b}} + \frac{1}{4} \mu_{\Lambda \Sigma} F^{\Lambda}_{\mu \nu} F^{\Sigma \mu \nu}
+ \frac{1}{4} \nu_{\Lambda \Sigma} F^{\Lambda}_{\mu \nu} (*F^{\Sigma})^{\mu \nu}
- {\rm g}^2 V (z, \ol{z})
\right\} \, , 
\end{align}
where we adopt the mostly plus signature in the four-dimensional spacetime metric.
From now on we abbreviate the gravitational Newton's constant $\kappa$ and the gauge coupling constant ${\rm g}$. 
The complex scalar field of the vector multiplet is given by $z^a$,  
whose index $a$ runs $a = 1,2,\dots,n_V$, 
whilst the index $\Lambda$ of $F_{\mu \nu}^{\Lambda}$ runs $\Lambda = 0,1,2,\dots, n_V$. 
The scalar field $z^a$ parametrizes the vector moduli space, i.e., the special K\"{a}hler geometry of local type, endowed with the K\"{a}hler metric $G_{a \ol{b}}(z, \ol{z})$.
The detail will be exhibited in later discussions.
The gauge field $A_{\mu}^{\Lambda}$ has a non-canonical kinetic term and a topological term which contain the scalar dependent functions $\mu_{\Lambda \Sigma}(z, \ol{z})$ and $\nu_{\Lambda \Sigma} (z, \ol{z})$, respectively. These two are given as
\begin{align}
\mu_{\Lambda \Sigma} (z, \ol{z}) \ &= \ {\rm Im} {\cal N}_{\Lambda \Sigma} (z, \ol{z})
\, , \ls
\nu_{\Lambda \Sigma} (z, \ol{z}) \ = \ {\rm Re} {\cal N}_{\Lambda \Sigma}(z, \ol{z})
\, , \label{munuN}
\end{align}
where ${\cal N}_{\Lambda \Sigma} (z, \ol{z})$ is a negative-definite function on the special K\"{a}hler geometry. The field strength of the gauge field and its Poincar\'{e} dual are 
\bsubeq
\begin{align}
F^{\Lambda}_{\mu \nu} \ &= \ 
\del_{\mu} A^{\Lambda}_{\nu} - \del_{\nu} A^{\Lambda}_{\mu} 
\, , \ls
(*F^{\Lambda})_{\mu \nu} \ = \ 
\frac{\sqrt{-g}}{2} \ve_{\mu \nu \rho \sigma} F^{\Lambda \rho \sigma}
\, , 
\end{align}
\esubeq
where we introduced the $\ve$-tensor defined as
$\ve_{0123} = 1$ and $\ve^{0123} = g^{-1}$.
Notice that the indices $\mu$, $\nu$, $\rho$ and $\sigma$ represent the curved spacetime coordinates\footnote{The ones in the orthogonal frame are described as $\wh{a}$ and the $\ve$-tensor in this frame is given as $\ve_{\wh{0}\wh{1}\wh{2}\wh{3}} = 1 = - \ve^{\wh{0}\wh{1}\wh{2}\wh{3}}$. In this paper, however, we do not the orthogonal frame of the spacetime, whilst the orthogonal frame of the vector moduli space will appear in section \ref{more_formal}.}.
The equations of motion 
and the Bianchi identity are 
\bsubeq \label{EOM}
\begin{align}
&R_{\mu \nu} - \half R g_{\mu \nu} 
- 2 G_{a \ol{b}} \del_{(\mu} z^a \del_{\nu)} \ol{z}{}^{\ol{b}}
+ g^{\rho \sigma} G_{a \ol{b}} \del_{\rho} z^a \del_{\sigma} \ol{z}{}^{\ol{b}} 
g_{\mu \nu} 
\ = \ 
T_{\mu \nu} - V g_{\mu \nu}
\, , \\
&T_{\mu \nu} \ = \ 
- \mu_{\Lambda \Sigma} F^{\Lambda}_{\mu \rho} F^{\Sigma}_{\nu \sigma} g^{\rho \sigma}
+ \frac{1}{4} \mu_{\Lambda \Sigma} F^{\Lambda}_{\rho \sigma} F^{\Sigma \rho \sigma} g_{\mu \nu} 
\, , \\
&\ve^{\mu \nu \rho \sigma} \del_{\nu} G_{\Lambda \rho \sigma} 
\ = \ 
0 
\, , \ls
G_{\Lambda \mu \nu} \ = \ 
- \mu_{\Lambda \Sigma} (*F^{\Sigma})_{\mu \nu}
+ \nu_{\Lambda \Sigma} F^{\Sigma}_{\mu \nu} 
\, , \\
&- \frac{G_{a \ol{b}}}{\sqrt{-g}} \del_{\mu} \big(
\sqrt{-g} g^{\mu \nu} \del_{\nu} \ol{z}{}^{\ol{b}} \big)
- g^{\rho \sigma} \frac{\del G_{a \ol{b}}}{\del \ol{z}{}^{\ol{c}}} 
\del_{\rho} \ol{z}{}^{\ol{b}} \del_{\sigma} \ol{z}{}^{\ol{c}}
\ = \ 
\frac{1}{4} \frac{\del \mu_{\Lambda \Sigma}}{\del z^a} F^{\Lambda}_{\mu \nu} F^{\Sigma \mu \nu} 
+ \frac{1}{4} \frac{\del \nu_{\Lambda \Sigma}}{\del z^a} F^{\Lambda}_{\mu \nu} (*F^{\Sigma})^{\mu \nu} 
- \frac{\del V}{\del z^a}
\, , \\
&\ve^{\mu \nu \rho \sigma} \del_{\nu} F^{\Lambda}_{\rho \sigma} 
\ = \ 0
\, .
\end{align}
\esubeq
Let us define electric charges $q_{\Lambda}$ and magnetic charges $p^{\Lambda}$ of the abelian gauge field. 
In terms of the field strength $F^{\Lambda}$ and its dual $G_{\Lambda}$ we define these two charges in such a way as
\begin{align}
q_{\Lambda} \ &= \ 
\frac{1}{4 \pi} \int_{S^2} G_{\Lambda}
\, , \ls
p^{\Lambda} \ = \ 
\frac{1}{4 \pi} \int_{S^2} F^{\Lambda}
\, .
\end{align}
They can be interpreted as charges of D-branes wrapped on certain cycles on the internal six-dimensional geometry in string theory compactification scenarios\footnote{In a usual black hole attractor we introduce a Calabi-Yau three-fold as the internal space.}: 
$(q_0, q_a, p^a, p^0)$ are the charges of D0, D2, D4 and D6-branes in type IIA string theory: the Ramond-Ramond two-form, related to D0-branes, yields the electric charge $q_0$; the four-forms corresponding to D2-branes wrapped on two-cycles give the charges $q_a$; the six- and eight-forms, corresponding to D4- and D6-branes wrapped on four- and six-cycles, generate the magnetic charges $p^a$ and $p^0$, respectively. 
In later discussions we often use this terminology by convention.

\subsection{Scalar potential}

In gauged supergravity the scalar potential emerges and plays an important role in the search of vacua. 
It also provides non-trivial configurations in the investigation of black hole solutions.
In the absence of the FI parameters,
the form $V(z, \ol{z})$ in the abelian gauged supergravity without hypermultiplets is described as \cite{Andrianopoli:1996cm, DallAgata:2003yr, DAuria:2004yi}
\begin{align}
V (z, \ol{z}) \ &= \ 
\Big(
U^{\Lambda \Sigma} - 3 \ol{L}{}^{\Lambda} L^{\Sigma} \Big) 
{\cal P}^x_{\Lambda} {\cal P}^x_{\Sigma} 
\nn \\
\ & \ \ \ \ 
+ \Big\{ G^{a \ol{b}} \big(
f_a^{\Lambda} h_{\Sigma \ol{b}} + f_{\ol{b}}^{\Lambda} h_{\Sigma a}
\big) 
- 3 \big(
\ol{M}_{\Lambda} L^{\Sigma} + M_{\Lambda} \ol{L}{}^{\Sigma} 
\big) \Big\}
{\cal P}^x_{\Sigma} \wt{\cal P}^{x \Lambda}
\nn \\
\ & \ \ \ \ 
+ \Big( G^{a \ol{b}} h_{\Lambda a} h_{\Sigma \ol{b}} - 3 \ol{M}_{\Lambda} M_{\Sigma} \Big) \wt{\cal P}^{x\Lambda} \wt{\cal P}^{x\Sigma} 
\, , 
\end{align}
where we used the property of the special K\"{a}hler geometry of local type which governs the moduli space of the vector multiplets\footnote{Notice that here we set all hypermultiplets and tensor multiplets to be zero.
Thus we can simply extract the terms of the (abelian) vector multiplets from the scalar potential (4.8) in \cite{DallAgata:2003yr}.}. 
The special K\"{a}hler geometry is represented in terms of the holomorphic sections $(X^{\Lambda}, {\cal F}_{\Lambda})$. Here we introduce various useful forms\footnote{If the prepotential ${\cal F}$ exists, this should be a homogeneous function of $X^{\Lambda}$ of degree two, and the section can be written as ${\cal F}_{\Lambda} = \del {\cal F}/\del X^{\Lambda}$. We assume that there exists an appropriate prepotential in each model in later discussions.}:  
\bsubeq \label{SKG}
\begin{align}
K \ &= \ - \log \big[ i (\ol{X}{}^{\Lambda} {\cal F}_{\Lambda} - X^{\Lambda} \ol{\cal F}_{\Lambda}) \big]
\, , \ls
z^a \ = \ \frac{X^a}{X^0} 
\, , \ls 
G_{a \ol{b}} \ = \ \frac{\del}{\del z^a} \frac{\del}{\del \ol{z}{}^{\ol{b}}} K 
\, , \label{SKG1} \\
\Pi \ &= \ \e^{K/2} \left(
\begin{array}{c}
X^{\Lambda} \\
{\cal F}_{\Lambda}
\end{array} \right)
\ = \ 
\left(
\begin{array}{c}
L^{\Lambda} \\
M_{\Lambda}
\end{array} \right)
\, , \ls
D_a \Pi \ = \ 
\Big( \frac{\del}{\del z^a} + \half \frac{\del K}{\del z^a} \Big) \Pi
\ = \ 
\left(
\begin{array}{c}
f_a^{\Lambda} \\
h_{\Lambda a}
\end{array} \right)
\, , \label{SKG2} \\
M_{\Lambda} \ &= \ {\cal N}_{\Lambda \Sigma} L^{\Sigma}
\, , \ls
h_{\Lambda a} \ = \ \ol{\cal N}_{\Lambda \Sigma} f_a^{\Sigma}
\, , \label{SKG3} \\
U^{\Lambda \Sigma} \ &= \ 
G^{a \ol{b}} f_a^{\Lambda} f_{\ol{b}}^{\Sigma}
\ = \ 
- \half {\rm Im}({\cal N}^{-1})^{\Lambda \Sigma} - \ol{L}{}^{\Lambda} L^{\Sigma}
\, . \label{SKG4}
\end{align}
\esubeq
In particular we refer to $K$ and $D_a$ as the K\"{a}hler potential and the K\"{a}hler covariant derivative (including the Levi-Civita connection), respectively. 
Two sets of variables $\{{\cal P}^x_{\Lambda}\}$ and $\{\wt{\cal P}^{x \Lambda} \}$ are called the triplets of Killing prepotentials ($x=1,2,3$) which play central roles in ${\cal N}=2$ gauged supergravity. These are represented in terms of scalar fields in the hypermultiplets and the vector multiplets (in a modern expression, see, for instance, \cite{Cassani:2009na}). 
Here we only consider the system with vector multiplets. 
In this case only the third Killing prepotential ${\cal P}^3 = 
({\cal P}^3_{\Lambda} L^{\Lambda} - \wt{\cal P}^{3 \Lambda} M_{\Lambda})$ contributes to the scalar potential $V$ in such a way as
\begin{align}
V \ &= \ 
G^{a \ol{b}} D_a {\cal P}^3 \ol{D_b {\cal P}^3}
- 3 |{\cal P}^3|^2
\, . \label{V_P3}
\end{align}

\subsection{Central charge} \label{centralcharge}

It is convenient to introduce the graviphoton charge, which is also referred to as the central charge $Z$, in later discussions. This is defined by the graviphoton field strength $T^{-}_{\mu \nu}$ appearing in the supersymmetry variations of gravitini
\cite{Andrianopoli:1996cm, DallAgata:2003yr, DAuria:2004yi}
\begin{align}
\delta \psi_{A \mu} \ &= \ 
D_{\mu} \ve_A + \epsilon_{AB} T^{-}_{\mu \nu} \gamma_{\mu} \ve^B 
+ i S_{AB} \gamma_{\mu} \ve^B
+ \dots
\, , \\
T^{-}_{\mu \nu} \ &= \ 
2 M_{\Lambda} F^{\Lambda}_{\mu \nu} - 2 L^{\Lambda} G_{\Lambda \mu \nu}
\, , \ls
S_{AB} \ = \ \frac{i}{2} (\sigma_x)_{AB} {\cal P}^x
\, , 
\end{align}
where $(\sigma_x)_{AB}$ is the Pauli matrix.
The central charge can be given in terms of the electric charges and the magnetic charges such as
\begin{align}
Z (z,\ol{z},p,q) \ &= \ 
- \half \left( \frac{1}{4 \pi} \int_{S^2} T^{-} \right)
\ = \ 
L^{\Lambda} q_{\Lambda} - M_{\Lambda} p^{\Lambda}
\nn \\
\ &= \ 
\e^{K/2} \big( X^{\Lambda} q_{\Lambda} - {\cal F}_{\Lambda} p^{\Lambda} \big)
\ = \ 
\e^{K/2} W (z, p,q)
\, .
\end{align}
We refer to $W(z, p,q)$ as the holomorphic central charge.
The K\"{a}hler covariant derivative of the central charge is 
$D_a Z = (\del_a + \half \del_a K) Z = \e^{K/2} (\del_a + \del_a K) W$, where $\del_a = \frac{\del}{\del z^a}$.
It is also instructive to introduce the first symplectic invariant \cite{Ceresole:1995ca}
\begin{align}
I_1 (z, \ol{z}, p, q) \ &= \ 
- \half \big( p^{\Lambda} \ q_{\Lambda} \big)
\left(
\begin{array}{cc}
\mu_{\Lambda \Sigma} + \nu_{\Lambda \Gamma} (\mu^{-1})^{\Gamma \Delta} \nu_{\Delta \Sigma} & 
- \nu_{\Lambda \Gamma} (\mu^{-1})^{\Gamma \Sigma} \\
- (\mu^{-1})^{\Lambda \Gamma} \nu_{\Gamma \Sigma} & (\mu^{-1})^{\Lambda \Sigma}
\end{array}
\right) 
\left(
\begin{array}{c}
p^{\Sigma} \\
q_{\Sigma}
\end{array} 
\right)
\nn \\
\ &= \ 
|Z|^2 + G^{a \ol{b}} D_a Z \ol{D_b Z}
\, . \label{VBH}
\end{align}
This function is non-negative since the function ${\cal N}_{\Lambda \Sigma}$ is negative definite and the K\"{a}hler metric $G_{a \ol{b}}$ is positive definite.
In the asymptotically flat black hole, this depicts the effective black hole potential $\VBH = V_{\text{BH}} (z, \ol{z}, p,q)$ \cite{Ferrara:1997tw}. 
In that spacetime the equation $\del_a \VBH = 0$ is called the black hole attractor equation. 
As we will see later, 
this is no longer the black hole attractor equation in the asymptotically AdS spacetime.

It is interesting if we can assign the Killing prepotentials ${\cal P}^3_{\Lambda}$ and $\wt{\cal P}^{3\Lambda}$ to the electric charges $q_{\Lambda}$ and the magnetic charges $p^{\Lambda}$, respectively. 
Indeed such an identification can be realized by the Ramond-Ramond flux charges via (non)geometric flux compactifications in ten-dimensional type II string theory \cite{Cassani:2009na}\footnote{If the hypermultiplets are introduced the Killing prepotential ${\cal P}^3$ further contains the universal hypermultiplet and the other Killing prepotentials ${\cal P}_{\pm} = {\cal P}_1 \pm i {\cal P}_2$ should also contribute to the scalar potential $V$. This is beyond the scape of this paper.}. 
In that case the scalar potential (\ref{V_P3}) is given in terms of the central charge in such a way as
\begin{align}
V \ &= \ G^{a \ol{b}} D_a Z \ol{D_b Z} - 3 |Z|^2
\, , \label{V_Z}
\end{align}
whose form also appears as the scalar potential $V_{{\cal N}=1} = G^{i \ol{j}} D_i {\cal Z} \ol{D_j {\cal Z}} - 3 |{\cal Z}|^2$ in four-dimensional ${\cal N}=1$ supergravity, where ${\cal Z}$ is not the central charge but related to the ${\cal N}=1$ superpotential ${\cal W}$ as ${\cal Z} = \e^{K/2} {\cal W}$.
Nevertheless we often refer to $\del_a V = 0$ as the ``flux vacua attractor'' equation in later discussions.


\section{Extremal RN-AdS black hole solutions}
\label{main}

\subsection{Metric ansatz: static, spherically symmetric configuration}
\label{metric_ansatz}

In the previous section we discussed the equations of motion and the scalar potential as well as the central charge. 
Their descriptions are generic in any spacetime configurations. 
From now on we focus on a certain configuration, i.e., an extremal, static, spherically symmetric, asymptotically flat or AdS, charged black hole.
This is referred to as the extremal RN(-AdS) black hole.
This spacetime configuration allows the following metric ansatz:
\bsubeq
\begin{align}
\d s^2 \ &= \ 
- \e^{2 A(r)} \d t^2 + \e^{2 B(r)} \d r^2 
+ \e^{2 C(r)} r^2 \big( \d \theta^2 + \sin^2 \theta \d \phi^2 \big)
\, , \\
\sqrt{-g} \ &= \ 
\e^{A+B+2C} r^2 \sin \theta
\, , \ls
\ve_{tr\theta\phi} \ = \ 1
\, , \ls 
\ve^{tr\theta\phi} \ = \ g^{-1} \ = \ 
- \frac{\e^{-2 (A+B+2C)}}{r^4 \sin^2 \theta}
\, .
\end{align}
\esubeq
Under this metric ansatz the gauge field strength can be simply written as
\begin{align}
F^{\Lambda}_{tr} \ &= \ 
\frac{\e^{A+B-2C}}{r^2} (\mu^{-1})^{\Lambda \Sigma} \big( q_{\Sigma} - \nu_{\Sigma \Gamma} p^{\Gamma} \big)
\, , \ls
F^{\Lambda}_{\theta \phi} \ = \ 
p^{\Lambda} \sin \theta
\, .
\end{align}
They appear in the energy-momentum tensor and the interactions of the scalar fields in the equations of motion (\ref{EOM}). 
We exhibit them in terms of the symplectic invariant (\ref{VBH}):
\bsubeq
\begin{gather}
T_t{}^t \ = \ T_r{}^r \ = \ - T_{\theta}{}^{\theta} \ = \ - T_{\phi}{}^{\phi}
\ = \ 
- \frac{\e^{-4C}}{r^4} \VBH
\, , \label{diag_EMT} \\
\frac{1}{4} \frac{\del \mu_{\Lambda \Sigma}}{\del z^a} F^{\Lambda}_{\mu \nu} F^{\Sigma \mu \nu} 
+ \frac{1}{4} \frac{\del \nu_{\Lambda \Sigma}}{\del z^a} F^{\Lambda}_{\mu \nu} (*F^{\Sigma})^{\mu \nu} 
\ = \ 
- \frac{\e^{-4C}}{r^4} \frac{\del \VBH}{\del z^a}
\, .
\end{gather}
\esubeq
Then the equations of motion provide the following a set of differential equations: 
\bsubeq \label{ODE} 
\begin{align}
&
-\e^{- 2 B} 
\Big[
\frac{1}{r^2} (1 - \e^{2 (B-C)}) + C' (3C' - 2B') 
+ \frac{2}{r} (3C' - B') 
+ 2 C''
\Big]
+ \e^{-2 B} G_{a \ol{b}} {z^{a}}{}' {\ol{z}^{\ol{b}}}{}'
\nn \\
&\LS \ = \ 
\frac{\e^{- 4C}}{r^4} \VBH + V
\, , \\
&
-\e^{-2 B} 
\Big[
\frac{1}{r^2} (1 - \e^{2(B-C)}) + C' (C' + 2 A') + \frac{2}{r} (C' + A')
\Big]
+ \e^{-2 B} G_{a \ol{b}} {z^{a}}{}' {\ol{z}^{\ol{b}}}{}'
\ = \ 
\frac{\e^{- 4C}}{r^4} \VBH + V
\, , \\
&
-\e^{-2 B} 
\Big[
A'' + C'' + A'(A'-B') + C'(A'-B'+C') + \frac{1}{r}(A'-B'+2C')
\Big]
- \e^{-2 B} G_{a \ol{b}} {z^{a}}{}' {\ol{z}^{\ol{b}}}{}'
\nn \\
&\LS \ = \ 
- \frac{\e^{- 4C}}{r^4} \VBH + V
\, , \\
&
\e^{-2 B} 
\Big[
G_{a \ol{b}} {\ol{z}{}^{\ol{b}}}{}''
+ \frac{\del G_{a \ol{b}}}{\del \ol{z}{}^{\ol{c}}}
{\ol{z}{}^{\ol{b}}}{}' {\ol{z}{}^{\ol{c}}}{}'
+ G_{a \ol{b}} {\ol{z}{}^{\ol{b}}}{}' \big( A'-B'+2C' + \frac{2}{r} \big) 
\Big]
\ = \
\frac{\e^{- 4C}}{r^4} \frac{\del \VBH}{\del z^a} + \frac{\del V}{\del z^a}
\, , 
\end{align}
\esubeq
where the prime means the derivative with respect to the radial coordinate $r$.

\subsection{Effective potential}
\label{effective_potential}

Let us focus on 
the near horizon geometry $\text{AdS}_2 \times S^2$ of the extremal RN-(AdS) black hole.
We introduce the following ansatz\footnote{It is possible to discuss AdS black holes with unusual topology \cite{Caldarelli:1998hg}. We do not discuss them in the present work.}:
\bsubeq \label{AdS_EH}
\begin{align}
A \ &= \ - B \ = \ \log \frac{r - \rH}{\rA}
\, , \ls
C \ = \ 
\log \frac{\rH}{r}
\, , \\
\d s^2 \ &= \ 
- \Big( \frac{r- \rH}{\rA} \Big)^2 \d t^2 
+ \Big( \frac{\rA}{r- \rH} \Big)^2 \d r^2 
+ \rH^2 \big( \d \theta^2 + \sin^2 \theta \d \phi^2 \big)
\nn \\
\ &= \ 
- \frac{\e^{2 \tau}}{\rA^2} \d t^2 + \rA^2 \d \tau^2 
+ \rH^2 \big( \d \theta^2 + \sin^2 \theta \d \phi^2 \big)
\, , 
\end{align}
\esubeq
where we defined a new radial coordinate $\tau = \log (r - \rH)$ in the last expression.
The event horizon is located at $\tau = - \infty$ 
where the original radial coordinate arrives at $r = \rH$. 
Note that $\rA$ and $\rH$ are the radii of $\text{AdS}_2$ and $S^2$, respectively. 
These two are described in terms of $V$ and $\VBH$ at the event horizon as we will see soon.
Suppose there exists an attractor mechanism in this configuration and we assume the derivatives ${z^a}{}'$ and ${z^a}''$ vanish at the event horizon $r = \rH$ \cite{Bellucci:2008cb}. 
We can easily reduce the equations of motion (\ref{ODE}) to 
\bsubeq
\begin{align}
\frac{1}{\rH^2} \ &= \ 
\frac{1}{\rH^4} \VBH + V \Big|_{r = \rH}
\, , \label{AdS_EH1} \\
\frac{1}{\rA^2} \ &= \ 
\frac{1}{\rH^4} \VBH - V \Big|_{r = \rH}
\, , \label{AdS_EH2} \\
0 \ &= \ 
\frac{1}{\rH^4} \frac{\del \VBH}{\del z^a} + \frac{\del V}{\del z^a} \Big|_{r = \rH}
\, . \label{AdS_EH3} 
\end{align}
\esubeq
At the event horizon the two variables $\rH$ and $\rA$ are given as
\begin{align}
\rH^2 \ &= \ 
\frac{1 - \sqrt{1 - 4 \VBH V}}{2 V} \Big|_{r = \rH}
\, , \ls
\rA^2 \ = \ 
\frac{\rH^2}{\sqrt{1 - 4 \VBH V}} \Big|_{r = \rH}
\, ,
\end{align}
where $1 - 4 \VBH V > 0$ is postulated.
The scalar curvature in four-dimensional spacetime is given as $R = 2 (\rH^{-2} - \rA^{-2}) = 4 V$. 
Then the scalar potential at the event horizon is interpreted as the cosmological constant $V |_{r = \rH} = \Lambda$. 
Here we set $\Lambda < 0$ in the case of the asymptotically $\text{AdS}_4$ spacetime\footnote{Other situations $\rA = \rH$ or $\rA > \rH$ describe the asymptotically flat or de Sitter four-dimensional spacetime, respectively. In these two case the scalar potential becomes $V = 0$ or $V > 0$ at the event horizon.}.
The third equation (\ref{AdS_EH3}) is rewritten as
\begin{align}
0 \ &= \ 
\frac{1}{\rH^4} \big( 1 - 2 \rH^2 V \big) \frac{\del}{\del z^a} 
\rH^2 \Big|_{r = \rH}
\, .
\end{align}
Here it is enough to consider the case when $\del \rH^2/\del z^a |_{r = \rH}$ vanishes, because the situation $\rH = \infty$ seems unnatural and the case $1 - 2 \rH^2 V = 0$ gives the positive $V$. 

The Bekenstein-Hawking entropy $S_{\text{BH}}$ corresponds to the area of the event horizon divided by $4\pi$, i.e., $S_{\text{BH}} = A_{\text{H}}/4 \pi = \rH^2$ \cite{Morales:2006gm}. 
In addition the entropy corresponds to the free-energy defined as the Legendre transformation of the effective action of the system. 
Then it is also useful to introduce the effective potential of the black hole near the event horizon even in the asymptotically AdS spacetime \cite{Bellucci:2008cb}
\begin{align}
\VE (z, \ol{z},p,q) \ &= \ 
\frac{1 - \sqrt{1 - 4 \VBH V}}{2 V}
\, , \label{VE}
\end{align}
where the effective potential coincides with the black hole entropy at the event horizon, which means $\VE|_{r = \rH} = \rH^2 = S_{\text{BH}}$.
In the asymptotically flat limit this function is smoothly connected to $\VBH$ since the function $\VE$ can be expanded in terms of $V$ such as \cite{Bellucci:2008cb}
\begin{align}
\VE 
\ &= \ 
\frac{1}{2V} + \half \sum_{n=0}^{\infty} 
\left( \begin{array}{c}
2n \\
n
\end{array} \right)
\frac{(\VBH)^n V^{n-1}}{2n-1}
\ = \ 
\VBH + (\VBH)^2 V + 2 (\VBH)^3 V^2 + {\cal O}((\VBH)^4V^3)
\, . \label{power_VE}
\end{align}
We have to evaluate the following equation in order to find the black hole solutions under the constraint $1 - 4 \VBH V > 0$:
\begin{align}
0 \ &= \ \frac{\del}{\del z^a} \VE \Big|_{r = \rH}
\nn \\
\ &= \ 
\frac{1}{2 V^2 \sqrt{1 - 4 \VBH V}} \Big\{
2 V^2 \frac{\del}{\del z^a} \VBH 
- \big( \sqrt{1 - 4 \VBH V} + 2 \VBH V - 1 \big) \frac{\del}{\del z^a} V
\Big\} \Big|_{r = \rH}
\, . \label{eq_VE} 
\end{align}
This equation still allows $V|_{r = \rH} = 0$ because the effective potential is expanded with respect to non-negative powers of $V$ as in (\ref{power_VE}). 
On the other hand, we impose that another factor $1 - 4 \VBH V$ in the denominator is positive definite.
Since the potentials $\VBH$ and $V$ are written in terms of the central charges $Z$ in such a way as (\ref{VBH}) and (\ref{V_Z}),
we can describe their derivatives as follows:
\bsubeq \label{attractor_eqs}
\begin{align}
\del_a \VBH \ &= \ 
2 \ol{Z} D_a Z + i C_{abc} G^{b \ol{b}} G^{c \ol{c}} \ol{D_b Z} \, \ol{D_c Z}
\, , \label{BH_attractor} \\
\del_a V \ &= \ 
-2 \ol{Z} D_a Z + i C_{abc} G^{b \ol{b}} G^{c \ol{c}} \ol{D_b Z} \, \ol{D_c Z}
\, , \label{FV_attractor}
\end{align}
\esubeq
where $C_{abc}$ is the totally symmetric K\"{a}hler-covariantly holomorphic tensor in the special K\"{a}hler geometry given as
$C_{abc} =
\e^K (\del_a X^{\Lambda} \del_b X^{\Sigma} \del_c X^{\Gamma}) \del_{\Lambda} \del_{\Sigma} \del_{\Gamma} {\cal F}$.
As mentioned in the end of section \ref{setup}, 
these two forms often appear in the ${\cal N}=2$ black hole attractors or in the ${\cal N}=1$ flux vacua attractors. 

\subsection{Search for extremal RN-AdS black hole solutions}
\label{search}

Now let us start to search an extremal RN-AdS black hole solution of the equation (\ref{eq_VE}) under the negative valued potential $V < 0$. 
There we often abbreviate the symbol ``$|_{r = \rH}$''.
Since this equation contains $\del_a \VBH$ and $\del_a V$ 
one might use various techniques developed in the black hole/flux vacua attractors. 
Unfortunately, however, the present situation is not so simple and 
possible solutions are completely different from the ones in the black hole/flux vacua attractors.
Here we explore possible types which satisfy the equation (\ref{eq_VE}) under the constraints $V < 0$ and $1 - 4 \VBH V > 0$:
\begin{enumerate}
\item \label{case1} $\del_a \VBH = 0$ and $\del_a V = 0$
\item \label{case2} $\del_a \VBH = 0$ and $\sqrt{1 - 4 \VBH V} + 2 \VBH V - 1 = 0$
\item \label{case5} $2 V^2 \del_a \VBH - (\sqrt{1 - 4 \VBH V} + 2 \VBH V - 1) \del_a V = 0$ with constraints $\del_a \VBH \neq 0$, $\del_a V \neq 0$ and $\sqrt{1 - 4 \VBH V} + 2 \VBH V -1 \neq 0$
\end{enumerate}
It is easy to analyze the former two types \ref{case1} and \ref{case2}, which will not generate any extremal RN-AdS black hole solutions with well-defined event horizon. 
It is rather difficult to find a solution in the last type \ref{case5}.

\subsubsection{Type \ref{case1}: strong attractor}

Here let us consider the type \ref{case1}. 
This is very restrictive because both of the forms in (\ref{attractor_eqs}) should vanish.
Immediately it turns out that $\ol{Z} D_a Z = 0$ and $i C_{abc} G^{b \ol{b}} G^{c \ol{c}} \ol{D_b Z} \, \ol{D_c Z} = 0$ should be realized simultaneously.
There exist three cases: (a) $Z \neq 0$ and $D_a Z = 0$ for any $D_a Z$; (b) $Z = 0$ and $D_a Z = 0$ for any $D_a Z$; (c) $Z = 0$ and $D_a Z \neq 0$ for a certain $D_a Z$ but $i C_{abc} G^{b \ol{b}} G^{c \ol{c}} \ol{D_b Z} \, \ol{D_c Z} = 0$.

The case (a) is similar to the case of the extremal BPS RN black holes in the asymptotically flat spacetime \cite{Ferrara:1995ih, Goldstein:2005hq}. 
But the physical situation is quite different.
Since the central charge does not vanish, 
all the covariant derivatives have to be zero.
This implies that a solution is supersymmetric and the scalar potential is strictly negative $V = - 3 |Z|^2$. 
This gives an extremal supersymmetric RN-AdS black hole. 
However, it is known that this configuration has a naked singularity in order to satisfy the vanishing supersymmetry variations of gravitini. 
The naked singularity emerges in the electric \cite{Romans:1991nq}, magnetic, 
and dyonic black holes \cite{Chamseddine:2000bk} in the four-dimensional asymptotically AdS spacetime. 
Thus we abandon this case when we consider a RN-AdS black hole with a regular event horizon.

The case (b) satisfies the equation (\ref{eq_VE}) trivially. This configuration gives the vanishing scalar potential $V = 0$ at the event horizon. In particular the radius of the event horizon $\rH$ also becomes zero. This is referred to as the empty hole (see, for instance, \cite{Denef:2000nb}) and is not considered as a RN-AdS black hole solution. 

The case (c) has also been well-developed in the asymptotically flat black holes \cite{Bellucci:2006xz}. 
This provides the non-negative scalar potential $V \geq 0$ 
because the central charge vanishes whilst the term $G^{a \ol{b}} D_a Z \ol{D_b Z}$ is positive semi-definite. Therefore this never gives RN-AdS black hole solutions.

We conclude that there never exist the extremal RN-AdS black hole solutions with well-defined event horizon in the type \ref{case1}.

\subsubsection{Type \ref{case2}: black hole attractor}

Next we consider the type \ref{case2}. 
Here we can also use the technique of the black hole attractor. 
However the second constraint $\sqrt{1 - 4 \VBH V} + 2 \VBH V - 1 = 0$ provides a strong condition
\begin{align}
\VBH V \ = \ 0
\, .
\end{align}
Since we look for a solution which satisfies $V < 0$, it turns out that $\VBH = 0$. 
The value $\VBH$ described in (\ref{VBH}) is positive semi-definite, 
then we immediately show that $Z = 0$ and $G^{a \ol{b}} D_a Z \ol{D_b Z} =0$ are realized simultaneously. 
This conflicts with the restriction $V < 0$. 
Then we find no suitable solutions of the extremal RN-AdS black hole.

\subsubsection{Type \ref{case5}: attractor equation for RN-AdS black hole}

Now let us study the type \ref{case5} which seems difficult to solve
under various strong constraints. Here we again enumerate them:
\bsubeq \label{RNAdS_constraints}
\begin{align}
0 \ &= \ 
2 V^2 \del_a \VBH - \big( \sqrt{1 - 4 \VBH V} + 2 \VBH V - 1 \big) \del_a V
\, , \label{const1} \\
0 \ &> \ V 
\, , \label{const2} \\
0 \ &< \ 
1 - 4 \VBH V 
\, , \label{const3} \\
0 \ &\neq \ \del_a \VBH 
\, , \label{const4} \\
0 \ &\neq \ \del_a V 
\, , \label{const5} \\
0 \ &\neq \ 
\sqrt{1 - 4 \VBH V} + 2 \VBH V - 1 
\, . \label{const6}
\end{align}
\esubeq
Because of the negative-valued condition (\ref{const2}) the central charge $Z$ must be non-zero.
The constraint (\ref{const6}) combined with (\ref{const2}) implies that $\VBH$ does not vanish.
In addition, we further impose $D_a Z \neq 0$ for some $D_a Z$,
because if all of the K\"{a}hler covariant derivatives vanish 
a supersymmetric solution emerges.
There the regular event horizon does not appear as in the case \ref{case1}-(a). 
In terms of expressions (\ref{VE}) and (\ref{attractor_eqs}) we rewrite the equation (\ref{const1}) to
\begin{align}
0 \ &= \ 
2 \GV \ol{Z} D_a Z 
+ i C_{abc} G^{b \ol{b}} G^{c \ol{c}} \ol{D_b Z} \, \ol{D_c Z}
\, , \ls
\GV \ = \ 
\frac{1 - \VE^2}{1 + \VE^2}
\, . \label{eq_VE2}
\end{align}
This form might be reduced to the black hole attractor equation $\del_a \VBH = 0$ if $\GV = 1$ (\ref{BH_attractor}) or to the ``flux vacua attractor'' equation $\del_a V = 0$ if $\GV = -1$ (\ref{FV_attractor}). 
But the function $\GV$ never smoothly arrives at $+ 1$ ($\VE \to 0$) nor at $-1$ ($\VE \to \infty$) caused by the above constraints.

The point $\GV = 0$ is rather special.
In this case the effective potential has to be $\VE = \pm 1$. 
We can choose only $\VE = + 1$ since the effective potential has to be identified with the positive-valued $\rH^2$.
In this case we have to search a solution which satisfies 
\bsubeq \label{GV=0case}
\begin{align}
i C_{abc} G^{b \ol{b}} G^{c \ol{c}} \ol{D_b Z} \, \ol{D_c Z} \ &= \ 0
\, . 
\end{align}
Let us analyze it more. 
The expression (\ref{VE}) gives $4 V (V + \VBH - 1) = 0$, 
then we have to find a solution of $V + \VBH = 1$ under the condition (\ref{const2}).
Via the descriptions (\ref{VBH}) and (\ref{V_Z}), we have a new relation
\begin{align}
- |Z|^2 + G^{a \ol{b}} D_a Z \ol{D_b Z} \ &= \ \half
\, .
\end{align}
\esubeq
Furthermore the condition (\ref{const3}) is rewritten as $(2 \VBH -1)^2 > 0$.
Then there would exist a non-supersymmetric solution if the symplectic invariant satisfies $\VBH = 1 - V > 1$.
Note that this is a strange solution in the physical sense. 
The black hole entropy given by the effective potential emerges as a constant $S_{\text{BH}} = \VE|_{r = \rH} = 1$ which does not depend on any black hole charges $(p^{\Lambda}, q_{\Lambda})$. 



\subsection{A formula}
\label{more_formal}

It seems that the following formula, 
appeared as in eq.(2.21) in \cite{Kallosh:2006ib},
must be still instructive among the charges $\Gamma = (p^{\Lambda} , q_{\Lambda})^\T$, the pair of coordinates $\Pi = (L^{\Lambda} , M_{\Lambda})^\T$, the central charge $Z$ and the symplectic invariant $\VBH$: 
\begin{align}
\Gamma^\T + i \frac{\del \VBH}{\del \wt{\Gamma}} 
\ =
2 i \ol{Z} \Pi^\T + 2 i G^{a \ol{b}} D_a Z \ol{D_b \Pi}^\T
\, . \label{Gamma_I}
\end{align}
This is apparently irrespective of the scalar potential $V$
even in the asymptotically non-flat spacetime.
Combining this with the attractor equation (\ref{eq_VE2}), 
we again encounter an intricate expression.
However it is quite important to reveal the most generic form of the solutions for future works.


First we consider the following matrix:
\bsubeq \label{matrix}
\begin{align}
&G^{a \ol{b}} D_a \Pi \otimes (\ol{D_b \Pi})^\T
\nn \\
&\LS \ = \ 
G^{a \ol{b}} \left(
\begin{array}{c}
D_a L^{\Lambda} \\
D_a M_{\Lambda}
\end{array}
\right)
\Big(
\ol{D_b L}{}^{\Sigma} \ \, \ol{D_b M}{}_{\Sigma}
\Big)
\ = \ 
G^{a \ol{b}} \left(
\begin{array}{cc}
D_a L^{\Lambda} \ol{D_b L}{}^{\Sigma} &
D_a L^{\Lambda} \ol{D_b M}{}_{\Sigma} \\
D_a M_{\Lambda} \ol{D_b L}{}^{\Sigma} &
D_a M_{\Lambda} \ol{D_b M}{}_{\Sigma} 
\end{array}
\right)
\nn \\
&\LS \ = \ 
G^{a \ol{b}} \left(
\begin{array}{cc}
- \half (\mu^{-1})^{\Lambda \Sigma} - \ol{L}{}^{\Lambda} L^{\Sigma}
&
- \half (\mu^{-1})^{\Lambda \Gamma} \nu_{\Gamma \Sigma} - \ol{L}{}^{\Lambda} M_{\Sigma} - \frac{i}{2} \\
- \half \nu_{\Lambda \Gamma} (\mu^{-1})^{\Gamma \Sigma} - \ol{M}_{\Lambda} L^{\Sigma} + \frac{i}{2} 
&
- \half \mu_{\Lambda \Sigma} - \half \nu_{\Lambda \Gamma} (\mu^{-1})^{\Gamma \Delta} \nu_{\Delta \Sigma} - \ol{M}_{\Lambda} M_{\Sigma}
\end{array}
\right)
\nn \\
&\LS \ = \ 
- \ol{\Pi} \otimes \Pi^\T
- \frac{i}{2} \left(
\begin{array}{cc}
0 & 1 \\
-1 & 0
\end{array}
\right)
- \half {\cal M}
\, , \\
&\ls \ \, {\cal M} \ = \ 
\left(
\begin{array}{cc}
(\mu^{-1})^{\Lambda \Sigma} 
&
(\mu^{-1})^{\Lambda \Gamma} \nu_{\Gamma \Sigma} \\
\nu_{\Lambda \Gamma} (\mu^{-1})^{\Gamma \Sigma} 
&
\mu_{\Lambda \Sigma} + \nu_{\Lambda \Gamma} (\mu^{-1})^{\Gamma \Delta} \nu_{\Delta \Sigma} 
\end{array} \right)
\, ,
\end{align}
\esubeq
where we used (\ref{SKG3}) and (\ref{SKG4}).
We also introduce the following forms:
\begin{align*}
\Gamma \ &= \ 
\left(\!
\begin{array}{c}
p^{\Lambda} \\
q_{\Lambda}
\end{array}
\!\right)
\, , \ls
\wt{\Gamma} \ = \ 
\left(
\begin{array}{cc}
0 & 1 \\
-1 & 0
\end{array}
\right)
\Gamma \ = \ 
\left(\!\!
\begin{array}{c}
q_{\Lambda} \\
-p^{\Lambda}
\end{array}
\!\!\right)
\, .
\end{align*}
Then the central charge and the first symplectic invariant can be written as
\begin{align*}
Z \ &= \ 
\wt{\Gamma}^\T \Pi
\, , \ls
\VBH \ = \ 
- \half \wt{\Gamma}{}^\T {\cal M} \wt{\Gamma}
\, .
\end{align*}
Dividing $\VBH$ by the charge $\wt{\Gamma}$ we see $\del \VBH/\del \wt{\Gamma} = - \wt{\Gamma}{}^\T {\cal M}$. 
Further we consider the acting $\wt{\Gamma}^\T$ on (\ref{matrix}) from the left:
\begin{align*}
G^{a \ol{b}} D_a Z \ol{D_b \Pi}{}^\T
\ &= \ 
- \ol{Z} \Pi^\T 
- \frac{i}{2} \Gamma^\T
- \half \wt{\Gamma}{}^\T {\cal M}
\ = \ 
- \ol{Z} \Pi^\T 
- \frac{i}{2} \Gamma^\T
+ \half \frac{\del \VBH}{\del \wt{\Gamma}}
\, .
\end{align*}
Then finally we obtain 
\begin{align*}
\Gamma^\T + i \frac{\del \VBH}{\del \wt{\Gamma}}
\ &= \ 
2 i \ol{Z} \Pi^{\T} + 2 i G^{a \ol{b}} D_a Z \ol{D_b \Pi}{}^{\T}
\, . 
\end{align*}
This is a generic formula which must satisfy in any configuration with any scalar potential.

\subsection{Examples}

\subsubsection{T$^3$-model}

We consider the T$^3$-model with a single vector multiplet. 
The prepotential ${\cal F}(X)$ is supposed to be cubic in the following way:
\begin{align}
{\cal F} \ &= \ \frac{(X^1)^3}{X^0}
\, , \ls
{\cal F}_0 \ = \ - \frac{(X^1)^3}{(X^0)^2} \ = \ - t^3 
\, , \ls
{\cal F}_1 \ = \ 3 \frac{(X^1)^2}{X^0} \ = \ 3 t^2
\, , \ls 
t \ = \ \frac{X^1}{X^0}
\, . \label{T3_data}
\end{align}
Introducing the indices in the orthogonal frame with hat symbol in the einbein $e_{\wh{1}}{}^t = i (t - \ol{t})/\sqrt{3}$, we rewrite the covariant derivatives:
\begin{align}
D_{\wh{1}} L^t \ &= \
e_{\wh{1}}{}^t D_t L^t \ = \ 
- \frac{i}{\sqrt{3}} \e^{K/2} \big( 2 t + \ol{t} \big)
\, , \ls
D_{\wh{1}} L^0 \ = \ 
- i \sqrt{3} \, \e^{K/2}
\, .
\end{align}
Since this is the single modulus model, there are no non-supersymmetric solutions which satisfy (\ref{GV=0case}) and we can analyze only in the case $\GV \neq 0$.
Suppose the central charge and its covariant derivative can be written as
$Z = - i \rho \e^{i(\alpha - 3 \phi)}$ and $D_{\wh{1}} Z = \sigma \e^{- i \phi}$ 
\cite{Kallosh:2006ib}, 
where $\rho$ and $\sigma$ are real positive values 
and $\alpha$ fixes the relative phase between $Z$ and $D_{\wh{1}} Z$. 
Actually this phase does not appear in the final result.
We further introduced a parameter $\phi$ which describes the K\"{a}hler transformation $K \to K + f(t) + \ol{f}(\ol{t})$, 
where $f(t)$ is an arbitrary holomorphic function of the modulus $t$. 
Under the K\"{a}hler transformation, the central charge and its covariant derivative are also transformed in such a way as $Z \to Z \e^{(\ol{f} - f)/2}$ and $D_{\wh{1}} Z \to D_{\wh{1}} Z \e^{(\ol{f} - f)/6}$.
Then one can relate the parameter $-i \phi$ to $(\ol{f} - f)/6$.
Since the left-hand side of (\ref{Gamma_I}) is manifestly K\"{a}hler invariant and in the right-hand side the phase cancels in each term separately.
Substituting them into the attractor equation (\ref{eq_VE2}) we find a relation between $\rho$ and $\sigma$:
\begin{align}
\sigma \ &= \ 
- \frac{\rho}{3} \e^{- i \alpha} \GV
\, .
\end{align}
Then the equation (\ref{Gamma_I}) provides a set of non-trivial equations among the black hole charges $(q_0, q, p, p^0)$, the symplectic invariant $\VBH$ and the modulus $t$ at the event horizon:
\bsubeq
\begin{align}
p + i \frac{\del \VBH}{\del q} 
\ &= \ 
- \frac{2 \rho}{3 \sqrt{3}} \e^{- i \alpha} \e^{K/2} 
\Big[ \big( 3 \sqrt{3} - 2 \GV \big) t - \GV \ol{t} \Big]
\, , \label{T3-pq_sol} 
\\
p^0 + i \frac{\del \VBH}{\del q_0} 
\ &= \ 
- \frac{2 \rho}{3} \e^{- i \alpha} \e^{K/2} \big( 3 - \sqrt{3} \GV \big)
\, . \label{T3-p0q0_sol}
\end{align}
\esubeq
Dividing (\ref{T3-pq_sol}) by (\ref{T3-p0q0_sol}), we obtain
\bsubeq
\begin{align}
\frac{p + i \frac{\del \VBH}{\del q}}{p^0 + i \frac{\del \VBH}{\del q_0}}
\ &= \ 
\frac{(3 \sqrt{3} - 2 \GV) t - \GV \ol{t}}{3 - \sqrt{3} \GV}
\, . \label{pqp0q0_t-T3}
\end{align}
Note that the symplectic invariant $\VBH$ is real (see, for instance, 
the diagonal components of the energy-momentum tensor (\ref{diag_EMT})). 
Then the complex conjugate of (\ref{pqp0q0_t-T3}) is simply
\begin{align}
\frac{p - i \frac{\del \VBH}{\del q}}{p^0 - i \frac{\del \VBH}{\del q_0}}
\ &= \ 
\frac{(3 \sqrt{3} - 2 \GV) \ol{t} - \GV t}{3 - \sqrt{3} \GV}
\, . \label{pqp0q0_tbar-T3}
\end{align}
\esubeq
Rewriting the above two forms we finally obtain a description of the modulus: 
\begin{align}
t \ &= \ 
\frac{3 \sqrt{3} - 2 \GV}{3 \sqrt{3} - \GV}
\left[
\frac{p + i \frac{\del \VBH}{\del q}}{p^0 + i \frac{\del \VBH}{\del q_0}}
\right]
+ \frac{\GV}{3 \sqrt{3} - \GV} 
\left[
\frac{p - i \frac{\del \VBH}{\del q}}{p^0 - i \frac{\del \VBH}{\del q_0}}
\right]
\, . \label{T3-formalsol}
\end{align}
This expression contains all the charges $(q_0, q, p, p^0)$. 
This is, however, highly nonlinear because the function $\GV$ appears in many places.
This function also contains the symplectic invariant $\VBH$ in such a way as (\ref{VE}).
Notice that we cannot take $\GV = 1$ (or $\VE = 0$) 
because this limit connects to the type \ref{case2} in section \ref{search}.
Even though it is hard to solve this description completely,
it may give rise to new intrinsic feature of the asymptotically non-flat black holes.

\subsubsection{STU-model}

Next we analyze the STU-model.
This is given by a cubic prepotential ${\cal F}(X) = X^1 X^2 X^3/X^0$.
We choose the local coordinates $z^a = X^a/X^0$.
The K\"{a}hler potential, the metric, the dribein and the totally symmetric tensor are
\bsubeq
\begin{gather}
K \ = \ 
- \log \big[ - i (z^1 - \ol{z}{}^1)(z^2 - \ol{z}{}^2)(z^3 - \ol{z}{}^3) \big]
\, , \\
G_{a \ol{b}} \ = \ 
- \frac{\delta_{ab}}{(z^a - \ol{z}{}^a)^2}
\, , \ls
e_{\wh{b}}{}^a \ = \ 
i \delta_{\wh{b}}^a (z^a - \ol{z}{}^a)
\, , \ls
C_{123} \ = \ \e^{K}
\, .
\end{gather}
\esubeq
The covariant derivatives which appear in (\ref{Gamma_I}) should be evaluated.
These are easily given as
\bsubeq
\begin{gather}
D_{\wh{b}} L^c \ = \ \sum_{a = 1}^3 e_{\wh{b}}{}^a D_a L^c 
\ = \ 
i \e^{K/2} \sum_{a=1}^3 \delta_{\wh{b}}^a \big\{ (z^a - \ol{z}{}^a) \delta_a^c - z^c
\big\}
\, , \ls
D_{\wh{b}} L^0 \ = \ 
-i \e^{K/2} \sum_{a=1}^3 \delta_{\wh{b}}^a 
\, , \\
\therefore \ \ \ 
i \sum_{\wh{b} =1}^3 \ol{D_{\wh{b}} L}{}^c \ = \ 
- \e^{K/2} (z^c + 2 \ol{z}{}^c)
\, , \ls
i \sum_{\wh{b}=1}^3 \ol{D_{\wh{b}} L}{}^0 \ = \ 
-3 \e^{K/2}
\, .
\end{gather}
\esubeq
Here we also suppose the central charge and its covariant derivatives have an isotropic configuration 
$Z = -i \rho \e^{i(\alpha - 3 \phi)}$ and $D_{\wh{a}}Z = \sigma \e^{- i \phi}$ 
\cite{Kallosh:2006ib}. 
It is natural to suppose the isotropy on the covariant derivatives via triality symmetry in this ${\cal N}=2$ configuration \cite{Duff:1995sm, Kallosh:2006ib}. 
In this case it is also natural to set $\GV \neq 0$. 
The parameters $\rho$, $\sigma$, $\alpha$, and $\phi$ work as in the T$^3$-model.
Substituting them into the attractor equation (\ref{eq_VE2}), the scale parameter $\sigma$ is given by 
\begin{align}
\sigma \ &= \ - \rho \e^{- i \alpha} \GV 
\, .
\end{align}
Then we can rewrite the equation (\ref{Gamma_I}) as follows:
\bsubeq
\begin{align}
p^a + i \frac{\del \VBH}{\del q_a} 
\ &= \ 
-2 \rho \e^{-i \alpha} \e^{K/2} \Big[
L^a + i \GV \sum_{\wh{b}=1}^3 \ol{D_{\wh{b}} L}{}^a \Big]
\ = \ 
-2 \rho \e^{-i \alpha} \e^{K/2} \Big[
(1 - \GV) z^a - 2 \GV \ol{z}{}^a
\Big]
\, , \label{STU-pq_sol} \\
p^0 + i \frac{\del \VBH}{\del q_0} 
\ &= \ 
-2 \rho \e^{-i \alpha} \e^{K/2} \Big[
L^0 + i \GV \sum_{\wh{b}=1}^3 \ol{D_{\wh{b}} L}{}^0 \Big]
\ = \ 
-2 \rho \e^{-i \alpha} \e^{K/2} (1 - 3 \GV)
\, . \label{STU-p0q0_sol}
\end{align}
\esubeq
Dividing (\ref{STU-pq_sol}) by (\ref{STU-p0q0_sol}), we obtain
\begin{align}
\frac{p^a + i \frac{\del \VBH}{\del q_a}}{p^0 + i \frac{\del \VBH}{\del q_0}}
\ &= \ 
\frac{(1 - \GV) z^a - 2 \GV \ol{z}{}^a}{1 - 3 \GV}
\; , \ls
\frac{p^a - i \frac{\del \VBH}{\del q_a}}{p^0 - i \frac{\del \VBH}{\del q_0}}
\ = \ 
\frac{(1 - \GV) \ol{z}{}^a - 2 \GV z^a}{1 - 3 \GV}
\, .
\end{align}
Then we eventually obtain the following form:
\begin{align}
z^a \ &= \ 
\frac{1 - \GV}{1 + \GV} \left[
\frac{p^a + i \frac{\del \VBH}{\del q_a}}{p^0 + i \frac{\del \VBH}{\del q_0}}
\right]
+ \frac{2 \GV}{1 + \GV} \left[
\frac{p^a - i \frac{\del \VBH}{\del q_a}}{p^0 - i \frac{\del \VBH}{\del q_0}}
\right]
\nn \\
\ &= \ 
\VE^2
\left[
\frac{p^a + i \frac{\del \VBH}{\del q_a}}{p^0 + i \frac{\del \VBH}{\del q_0}}
\right]
+ \big( 1 - \VE^2 \big)
\left[
\frac{p^a - i \frac{\del \VBH}{\del q_a}}{p^0 - i \frac{\del \VBH}{\del q_0}}
\right]
\, . \label{STU-formalsol}
\end{align}
This form contains all of the black hole charges $(p^{\Lambda}, q_{\Lambda})$. 
This is nonlinear because the function $\VE$ carries $\VBH$ (\ref{VE}).
Notice that we cannot take $\VE = 1$ unless $\GV = 0$.
We cannot also take $\VE = 0$ 
because this connects to the type \ref{case2} in section \ref{search}.
It indicates that the above solution is never reduced to neither BPS nor non-BPS ones in the black hole attractors in the asymptotically flat spacetime \cite{Kallosh:2006ib}.
Although it is hard to solve this description completely,
it may predict new features of the asymptotically non-flat black holes.

\section{Examples in T$^3$-model} \label{examples}

In this section we compute the attractor equation (\ref{eq_VE2}) and find a couple of concrete solutions in the T$^3$-model given by the cubic prepotential (\ref{T3_data}).
Since the K\"{a}hler potential $K(t, \ol{t})$, the totally symmetric covariant tensor $C_{abc}$ and the holomorphic central charge $W(t, p,q) = \e^{- K/2} Z(t, \ol{t}, p,q)$ are defined by the prepotential, they are also described as
\begin{gather}
\e^K \ = \ \frac{i}{(t- \ol{t})^3}
\, , \ls
\del_t K \ = \ - \frac{3}{t - \ol{t}}
\, , \ls
G_{t \ol{t}} \ = \ - \frac{3}{(t - \ol{t})^2}
\, , \ls
C_{ttt} \ = \ \frac{6i}{(t - \ol{t})^3}
\, , \\
W (t, p,q) \ = \ 
q_0 + q t - 3 p t^2 + p^0 t^3
\, . \label{T3_data2}
\end{gather}
As mentioned in section \ref{centralcharge}, 
we can interpret the electric charges and the magnetic charges as the ones of D-branes wrapped on certain cycles on the internal space in type IIA string theory. 
In the next discussions we will analyze the T$^3$-model under certain distributions of the D-brane charges; i.e., 
we focus on two typical cases, the D0-D4 system and the D2-D6 system.

\subsection{Magnetic configuration: D0-D4 black hole system}

First we discuss a black hole called the magnetic configuration. 
This black hole is dressed only by the charges $q_0$ and $p$. 
The holomorphic central charge $W(t,p,q)$ is simply reduced to
\begin{align}
W_{04} (t, p,q) \ &= \ 
q_0 - 3 p t^2 
\, , \label{W_D0D4}
\end{align}
whose discriminant is given as $\Delta (W_{04}) = 12 p q_0$.
It is rather difficult to find a generic solution.
Notice that the K\"{a}hler potential depends only on the imaginary part of $t$.
Here we look for a solution which can be given as $t = 0 + i y$ where $y \in {\mathbb R}$.
In order for the K\"{a}hler potential $K$ to be well-defined, the value $y$ should be strictly negative.
Introducing a variable $Y = (-y)^2$, we can represent the solution of 
(\ref{eq_VE2}) in the following algebraic description:
\begin{align}
f_{04} (Y) \ &= \ 
p Y^3 + (q_0 - 18 p^3 q_0^2) Y^2 - 12 p^2 q_0^3 Y - 2 p q_0^4
\ = \ 0
\, .
\end{align}
We find that both the charges $p$ and $q_0$ do not vanish in order to obtain a non-trivial solution $y < 0$.
The central charge and its covariant derivative are fixed as
\begin{align}
Z \Big|_{r = \rH} 
\ &= \ 
\frac{q_0 + 3 p y^2}{2} \sqrt{- \frac{1}{2y^3}} 
\, , \ls
D_t Z \Big|_{r = \rH} 
\ = \
\frac{3i (q_0 - p y^2)}{4 y} \sqrt{- \frac{1}{2y^3}} 
\, .
\end{align}
Here it is useful to consider the derivative $g_{04}(Y) = \del f_{04}/\del Y$ and the discriminants of the functions $f_{04}(Y)$ and $g_{04}(Y)$:
\bsubeq
\begin{align}
g_{04}(Y) \ &= \ 
3p Y^2 + 2 (q_0 - 18 p^3 q_0^2) Y - 12 p^2 q_0^3
\, , \\
\Delta (f_{04}) \ &= \ 
\frac{4 q_0^6}{p^2} \big( 2 p^3 q_0 + 9 (p^3 q_0)^2 + 432 (p^3 q_0)^3 \big)
\nn \\
\ &= \ 
\frac{(p q_0) q_0^2}{2}
\Big[ \big( 4 q_0^2 + 9(p q_0)^3 \big)^2 + 3375 (pq_0)^6 \Big] 
\, , \\
\Delta (g_{04}) \ &= \ 
4 q_0^2 \Big[ 1 + (18 p^3 q_0)^2 \Big]
\, .
\end{align}
\esubeq
The discriminant $\Delta (g_{04})$ is strictly positive. 
A formal solution $Y = (-y)^2>0$ is written as
\bsubeq
\begin{align}
Y \ &= \ 
\{ A_{04} + B_{04} \, , A_{04} + \omega^{\pm 1} B_{04} \}
\, \ls \text{with} \ \ \ \omega^3 \ = \ 1
\, , \\
A_{04} \ &= \ 
\frac{q_0}{3p} \big( 18 p^3 q_0 - 1 \big)
\, , \ls
B_{04} \ = \ 
\frac{1}{3p} \left(
C_{04}^{1/3} 
+ \frac{q_0^2 \{1 + (18 p^3 q_0)^2\}}{C_{04}^{1/3}} 
\right)
\, , \\
C_{04} \ &= \ 
- q_0^3 \Big[ 
1 - 27 p^3 q_0 - (18 p^3 q_0)^3 - 3 \sqrt{3} 
\sqrt{-2 p^3 q_0 - 9 (p^3 q_0)^2 - 432 (p^3 q_0)^3}
\Big]
\, .
\end{align}
\esubeq
In order to obtain a real positive solution $Y$, we have to impose $\Delta (f_{04}) < 0$ with a constraint $p q_0 <0$.
Substituting the appropriate values of $A$, $B$ and $C$ into $y = - \sqrt{Y}$, 
we obtain the symplectic invariant $I_1 |_{r = \rH}$, the cosmological constant $\Lambda = V |_{r = \rH}$ and the black hole entropy $S_{\text{BH}} = \VE |_{r = \rH}$:
\bsubeq \label{sol_D0D4}
\begin{align}
\VBH \big|_{r=\rH} \ &= \ 
|Z|^2 + G^{t \ol{t}} D_t Z \ol{D_t Z} \Big|_{r = \rH} 
\ = \ 
- \frac{q_0^2 + 3 p^2 y^4}{2 y^3} \ > \ 0
\, , \\
V \big|_{r=\rH} \ &= \ 
-3 |Z|^2 + G^{t \ol{t}} D_t Z \ol{D_t Z} \Big|_{r = \rH} 
\ = \ 
\frac{3p (q_0 + p y^2)}{y} 
\ = \ 
\frac{6(pq_0)^2 (q_0 + 3p y^2)^2}{y^5}
\ < \ 0
\, , \\
\VE \big|_{r=\rH} \ &= \ 
\frac{1 - \sqrt{1 - 4 \VBH V}}{2 V} \Big|_{r = \rH} 
\nn \\
\ &= \ 
\frac{- y}{12 (p q_0)^2 (q_0 + 3p y^2)^2}
\left\{ - y^4 + \sqrt{
y^8 + 12 (p q_0)^2 (q_0 + 3p y^2)^2 (q_0^2 + 3p^2 y^4)
}
\right\}
\ > \ 0
\, .
\end{align}
\esubeq
We remark that these values do not correspond to the ones of the non-BPS RN black hole solutions in the asymptotically flat spacetime because we imposed that the function $V$ is negative definite. This indicates that the above solution is isolated to any asymptotically flat black hole solutions. This is a new solution.
In this analysis, 
it is difficult to write down (\ref{sol_D0D4}) without using $y$ because of the intricate description of the solution $Y$. 
Moreover, since we analyzed the solution which must follow the conditions $Z \neq 0$ and $D_t Z \neq 0$, the value $\VE|_{r=\rH}$ would be endowed with not only the discriminants $\Delta(W_{04})$, $\Delta(f_{04})$ and $\Delta(g_{04})$ but also a more complicated structure.
It is known that, in the asymptotically flat extremal RN black hole case,  
the black hole entropy corresponds to the symplectic invariant $\VBH |_{r = \rH}$.
This can be given in terms of the discriminant (in the T$^3$-model) or, more generically, of the Cayley's hyperdeterminant (in the STU-model) 
even if both the central charge and the covariant derivative do not vanish \cite{Duff:2006uz, Bellucci:2007zi}.
In the asymptotically AdS case, however, the black hole entropy $S_{\text{BH}} = \VE|_{r = \rH}$ differs from the symplectic invariant $\VBH |_{r = \rH}$.
This is caused by the existence of the nonlinear function $\GV$ in the attractor equation (\ref{eq_VE2}).
This is not a constant but rather a complicated function of the modulus $t$. 

In order to argue asymptotic behaviors of the above solution,
let us discuss the large $|q_0|$ limit and the small $|q_0|$ limit.
For simplicity we look at the region $q_0 < 0$ with keeping $p q_0 < 0$.
In the small $|q_0|$ limit, the dominant term in each value is given as
\bsubeq
\begin{align}
y \big|_{r = \rH} \ &\sim \ 
- \sqrt{- \frac{2 q_0}{3 p}}
\, , \\
Z \Big|_{r = \rH} \ &\sim \ 
\frac{1}{4} \Big( - \frac{27}{2} p^3 q_0 \Big)^{\frac{1}{4}} 
\, , &
D_t Z \Big|_{r = \rH} \ &\sim \ 
\frac{15i}{8} p \Big( - \frac{3p}{8q_0} \Big)^{\frac{1}{4}}
\, , \\
\VBH \Big|_{r = \rH} \ &\sim \ 
\frac{7}{4} \sqrt{- \frac{3 p^3 q_0}{2}}
\, , &
\Lambda \ = \ V \Big|_{r = \rH} \ &\sim \ 
- \frac{27}{2} \sqrt{- \frac{3 (p^3 q_0)^3}{2}}
\, , \\
S_{\text{BH}} \ = \ \VE \Big|_{r = \rH} \ &\sim \ 
\frac{7}{4} \sqrt{- \frac{3 p^3 q_0}{2}}
\, .
\end{align}
\esubeq
These are quite similar to the ones in the non-BPS solution of the RN black hole in the asymptotically flat spacetime (see (\ref{info_D0D4_nonBPS}) in appendix \ref{non-BPS_RN}). Only one difference is the existence of non-vanishing cosmological constant $\Lambda = V |_{r = \rH}$ which is very small compared to $\VE|_{r = \rH} = S_{\text{BH}}$.
On the other hand, however, if $|q_0|$ is very large, 
the dominant term behaves in the following way:
\bsubeq
\begin{align}
y \big|_{r = \rH} \ &\sim \ 
\sqrt{30} \, p q_0
\, , \\
Z \Big|_{r = \rH} \ &\sim \ 
\frac{3}{2} \Big( \frac{15}{2} \Big)^{\frac{1}{4}} \sqrt{- p^3 q_0}
\, , &
D_t Z \Big|_{r = \rH} \ &\sim \ 
- \frac{i}{4} \Big( \frac{27}{40} \Big)^{\frac{1}{4}} \frac{1}{p q_0} \sqrt{- p^3 q_0}
\, , \\
\VBH \Big|_{r = \rH} \ &\sim \ 
- 3 \sqrt{\frac{15}{2}} (p^3 q_0) 
\, , &
\Lambda \ = \ V \Big|_{r = \rH} \ &\sim \ 
9 \sqrt{\frac{6}{5}} (p^3 q_0) 
\, , \\
S_{\text{BH}} \ = \ \VE \Big|_{r = \rH} \ &\sim \ 
\sqrt{\frac{5}{6}}
\, .
\end{align}
\esubeq
In this limit, the cosmological constant increases in the same power as the symplectic invariant $\VBH$, which originally yields the RN black hole entropy in the asymptotically flat spacetime. 
On the contrary, the effective potential $\VE$ is reduced to a constant which is independent of any black hole charges. 
This phenomenon is quite different from the black hole system which have been studied in the literature.

\subsection{D2-D6 black hole system}

Second we analyze another black hole solution in the D2-D6 system.
The holomorphic central charge $W(t,p,q)$ is reduced to
\begin{align}
W_{26} (t, p,q) \ &= \ 
q t + p^0 t^3
\, ,
\end{align}
whose discriminant is $\Delta (W_{26}) = -4 p^0 q^3$.
Here we also look for 
a solution which has only the imaginary part, i.e., $t = 0 + iy$ where $y \in {\mathbb R}$.
In order that the K\"{a}hler potential $K$ is well-defined, the value $y$ should be strictly negative.
Introducing a variable $Y = (-y)^2$, we can represent the solution of 
the attractor equation (\ref{eq_VE2}) in the following algebraic form:
\begin{align}
f_{26} (Y) \ &= \ 
2 (p^0)^4 q Y^3 -4 (p^0)^3 q^2 Y^2 + p^0 (3 + 2 p^0 q^3) Y - q
\ = \ 0
\, .
\end{align}
We find that both the charges $p^0$ and $q$ do not vanish in order to obtain a non-trivial solution $y < 0$.
At the event horizon the central charge and its covariant derivative are written as
\begin{align}
Z \Big|_{r = \rH} \ &= \ 
- i \big(q - p^0 y^2 \big) \sqrt{ - \frac{1}{8 y}}
\, , \ls
D_t Z \Big|_{r = \rH} \ = \ 
- \big( q + 3 p^0 y^2 \big) \sqrt{- \frac{1}{32 y^3}}
\, .
\end{align}
Here it is also useful to consider the derivative $g_{26}(Y) = \del f_{26}/\del Y$ and the discriminants:
\bsubeq
\begin{align}
g_{26}(Y) \ &= \ 
6 (p^0)^4 q Y^2 - 8 (p^0)^3 q^2 Y + p^0(3 + 2 p^0 q^3)
\, , \\
\Delta (f_{26}) \ &= \ 
- \frac{4 (p^0)^6}{q^2} \big( 54 p^0 q^3 - 9 (p^0 q^3)^2 + 16 (p^0 q^3)^3   \big) 
\nn \\
\ &= \ 
- (p^0q^3) \Big( \frac{(p^0)^3}{q} \Big)^2 \left[
\Big( 8 (p^0 q^3)^2 - \frac{9}{4} \Big)^2 
+ 3375 
\right]
\, , \\
\Delta (g_{26}) \ &= \ 
- 4 \frac{(p^0)^4}{q^2} \Big[ 18 p^0 q^3 -4 (p^0 q^3)^2 \Big]
\, .
\end{align}
\esubeq
Different from the D0-D4 system, the sign of the discriminant $\Delta (g_{26})$ is not restricted.
The formal description $Y = (-y)^2>0$ is written as
\bsubeq
\begin{align}
Y \ &= \ 
\{ A_{26} + B_{26} \, , A_{26} + \omega^{\pm 1} B_{26} \}
\, \ls \text{with} \ \ \ \omega^3 \ = \ 1
\, , \\
A_{26} \ &= \ 
\frac{2 q}{3 p^0}
\, , \ls
B_{26} \ = \ 
\frac{1}{6 (p^0)^2 q} 
\left(
C_{26}^{1/3} 
+ \frac{18 p^0 q^3 - 4 (p^0 q^3)^2}{q^2 C_{26}^{1/3}} 
\right)
\, , \\
C_{26} \ &= \ 
-p^0 \Big[
54 p^0 q^3 + 8 (p^0 q^3)^2 
- \frac{3\sqrt{3}}{p^0} \sqrt{(p^0)^2 (54 p^0 q^3 - 9 (p^0 q^3)^2 + 16 (p^0 q^3)^3)}
\Big]
\, .
\end{align}
\esubeq
It turns out that the discriminant $\Delta (f_{26})$ should be negative with a constraint $p^0 q^3 > 0$ in order for $C_{26}$ to be well-defined.
Substituting the appropriate values of $A$, $B$ and $C$ into $y = - \sqrt{Y}$, we can implicitly evaluate the following potentials at the event horizon $r = \rH$:
\bsubeq
\begin{align}
\VBH \big|_{r=\rH} 
\ &= \ 
|Z|^2 + G^{t \ol{t}} D_t Z \ol{D_t Z} \Big|_{r = \rH} 
\ = \ 
- \frac{q^2 + 3 (p^0)^2 y^4}{6 y} \ > \ 0
\, , \\
V \big|_{r=\rH} \ &= \ 
-3 |Z|^2 + G^{t \ol{t}} D_t Z \ol{D_t Z} \Big|_{r = \rH} 
\ = \ 
\frac{q(q -3 p^0 y^2)}{3y} 
\ = \ 
\frac{2 q^2 y}{3} \big( (p^0)^2 y^2 - q \big)^2  
\ < \ 0
\, , \\
\VE \big|_{r=\rH} \ &= \ 
\frac{1 - \sqrt{1 - 4 \VBH V}}{2V} \Big|_{r = \rH} 
\ = \
\frac{-3 + \sqrt{
9 + 4 q^2 \big( q - (p^0)^2y^2 \big)^2 \big( q^2 + 3 (p^0)^2 y^4 \big) 
}}{- 4 q^2 (q - (p^0)^2y^2)^2 y} 
\ > \ 0
\, .
\end{align}
\esubeq
It is confirmed 
that the symplectic invariant $I_1|_{r=\rH}$ and the black hole entropy $S_{\text{BH}} = \VE|_{r=\rH}$ are positive, whilst the cosmological constant $\Lambda = V|_{r=\rH}$ is strictly negative.
This solution also does not approach to non-BPS extremal RN black hole solutions in the asymptotically flat spacetime since we impose the negative definiteness on the scalar potential. This indicates that the above solution is isolated to the ones in the asymptotically flat spacetime. This is also a new solution.
Moreover, since we analyzed the solution which must follow the conditions $Z \neq 0$ and $D_t Z \neq 0$, the value $\VE|_{r=\rH}$ cannot be given simply in terms of the set of three discriminants $\Delta(W_{26})$, $\Delta(f_{26})$ and $\Delta(g_{26})$.
Here we omit the comparison between the above solution and the non-BPS solution of the RN black hole in the asymptotically flat spacetime.



\section{Discussions}
\label{discussions}

In the present work we searched the existence of non-supersymmetric solutions of the extremal RN-AdS black hole. 
Since it is not guaranteed that the negative cosmological constant emerges in the gauged supergravity without the FI parameters, first we discussed the existence of solutions satisfying the equations of motion. 
It turns out that such solutions do not appear if either the black hole attractor equation $\del_a \VBH = 0$ or the ``flux vacua attractor'' equation $\del_a V = 0$ is postulated. 
Despite of various strong constraints 
we obtained the new attractor equation (\ref{eq_VE2}) which differs from these attractor equations.
It is generically hard to solve this caused by the nonlinear function $\GV$ in front of the first term in (\ref{eq_VE2}).
Then we argued the formula (\ref{Gamma_I}) which does not depend on the scalar potential $V$. 
We described the modulus of non-supersymmetric solutions in the T$^3$-model (\ref{T3-formalsol}) and the ones in the STU-model (\ref{STU-formalsol}).
These descriptions are indeed general and nonlinear since the non-vanishing function $\GV$ (or the non-trivial form of $\VE$) contributes to the attractor equation (\ref{eq_VE2}) in these two models.
It would be interesting to elucidate their structures completely.

Next we analyzed the T$^3$-model with a cubic prepotential and found non-trivial solutions both in the D0-D4 black hole system and in the D2-D6 black hole system in a heuristic way.
The algebraic structures of the solutions contain the discriminants of the functions $\Delta(W)$, $\Delta (f)$ and $\Delta(g)$ which would govern the attractor equation (\ref{eq_VE2}). 
In the D0-D4 system, we further studied its asymptotic behaviors in the small and large $|q_0|$ limits. 
In the small $|q_0|$ limit, the central charge $Z$, the symplectic invariant $\VBH$, and the black hole entropy $S_{\text{BH}} = \VE |_{r = \rH}$ are quite similar to the ones in the non-BPS RN black hole solution.
There we also found that the non-vanishing cosmological constant $|\Lambda|$ is much smaller than the black hole potential $\VE$. 
In the large $|q_0|$ limit, however, the magnitude of the cosmological constant becomes very large whilst the black hole potential approaches to a constant which is independent of any black hole charges. 
It seems quite strange because the physical meaning of the increasing black hole charges indicates that many pieces of physical information fall down into the black hole and the black hole grows up. 
If this interpretation is correct, the black hole horizon should be larger and larger.
In order to realize this interpretation, 
we have to ``normalize'' the cosmological constant to a certain value and rescale the other physical values.
In this normalization one would find that the black hole horizon becomes larger and larger when the physical information falls into the black hole. 
But it is unclear the ``physical meaning of the normalization'' of various values.
This phenomenon quite differs from the case in the asymptotically flat extremal black holes.

One might anticipate that an appropriate solution exists if the discriminant of the holomorphic central charge $\Delta (W)$ is negative, as in the same analogy of the non-BPS RN black hole in the asymptotically flat spacetime or as in the search of flux vacua \cite{Kimura:2008tq}.
Unfortunately, however, as far as the value of the modulus $t$ in the T$^3$-model at the event horizon is restricted to be purely imaginary, this would be incorrect. 
One can immediately show that,
as far as we restrict the modulus $t$ to be purely imaginary, there are no solutions 
in certain charge distributions such as $( q_0, q, p, p^0) = (q_0, 0,0,p^0)$, $(0,q,p, 0)$, $(1,1,0,1)$ or $(1,0,-1,1)$ even though in all these cases the discriminants $\Delta (W)$ are strictly negative.
In the asymptotically flat black holes, we can (always) find non-BPS well-defined solutions even if we restrict the values of moduli to be purely imaginary. 
This also denotes a peculiarity of the RN-AdS black hole.
It would be interesting if this argument is resolved in the generic description (\ref{T3-formalsol}) which is rather complicated to extract the contribution of the discriminant $\Delta (W)$ in the present stage.

In this work we did not argue stability of the RN-AdS black holes which we obtained.
The effective potential (\ref{VE}) contains higher orders in $\VBH$ and $V$.
Then the Hessian matrix given by the second derivatives of $\VE$, which denotes mass eigenvalues of the moduli, would be highly complicated.
In general, non-supersymmetric solutions even in the asymptotically flat black holes, possess some flat directions. 
In order to check (semi-)stability of the black holes, one has to evaluate the third and fourth derivatives of the effective potential \cite{Nampuri:2007gv}.
In the presence of the FI parameters the (semi-)stability was already
discussed in \cite{Bellucci:2008cb}.
In the case without the FI parameters, it has not been well investigated yet. 
We will leave this topic in the present framework for future research.
{\sl Note added:} Recently it was reported that the stability of the extremal non-supersymmetric RN-AdS black holes in the context of charged membrane backgrounds of M-theory on $\text{AdS}_4 \times Y^7$ where $Y^7$ is a Sasaki-Einstein manifold
\cite{Klebanov:2010tj}.

It would be natural to introduce charged hypermultiplets \cite{Hristov:2010eu} 
as well as non-abelian vector multiplets in the context of compactifications in string theory.
In the framework of (non)geometric flux compactification scenarios
the scalar potential $V$ and its derivatives with respect to both the vector multiplets and the hypermultiplets were analyzed \cite{Cassani:2009na}, where the AdS vacua were discussed in ${\cal N}=0$ and ${\cal N}=1$ solutions. 

It is well-known that there exist supersymmetric solutions of asymptotically non-flat black holes with non-trivial angular momentum in the presence of well-defined event horizon (for instance, see \cite{Caldarelli:1998hg} and references therein). 
It would also be notable to search such a solution in four-dimensional ${\cal N}=2$ gauged supergravity with vector multiplets and hypermultiplets in the presence (or absence) of FI parameters. 
If a solution is directly derived from the string compactification,
one could acquire new viable prospects of nongeometric flux backgrounds 
which intertwines (non-)abelian gauge symmetries with string dualities via properties of black holes.


\section*{Acknowledgements}

The author would like to thank  
Davide Cassani,
Yoshinori Honma,
Wei Li,
Masahiro Ohta,
Susumu Okazawa 
and
Shinya Tomizawa
for valuable discussions.
He is also grateful to 
Kiril Hristov,
Yoshifumi Hyakutake,
Akihiro Ishibashi 
and 
Igor Klebanov
for useful comments.


\begin{appendix}

\section*{Appendix}

\section{Non-BPS solutions in asymptotically flat spacetime}
\label{non-BPS_RN}

In this appendix we briefly clarify a non-BPS RN black hole solution in the asymptotically flat spacetime. 
This indicates that we consider the configuration in which there are no hypermultiplets and the scalar potential vanishes $V = 0$. 
This coincides with the configuration in ${\cal N}=2$ ungauged supergravity.
For simplicity we just focus on the D0-D4 black hole system in the T$^3$-model.  

\subsection{D0-D4 system}

In the absence of the scalar potential $V$ in the system, 
the equation of motion (\ref{AdS_EH3}) is simply reduced to $0 = \del_a \VBH$ if the event horizon has a regular radius. 
This is nothing but the RN black hole attractor equation in the asymptotically flat spacetime. This is described as (\ref{BH_attractor}) in terms of the central charge.
In the T$^3$-model, all the information is given in (\ref{T3_data2}).
The central charge in the D0-D4 system is given in terms of (\ref{W_D0D4}) as
\begin{align}
Z \ &= \ \e^{K/2} W \ = \ 
\e^{K/2}
\big( q_0 - 3 p t^2 \big)
\, , \ls
D_t Z \ = \ 
\frac{3 \e^{K/2}}{t - \ol{t}} \big( p t^2 + 2 p t \ol{t} - q_0 \big)
\, .
\end{align}
In the search of an attractor point $t = 0 + i y$, we describe the attractor equation (\ref{BH_attractor}) as
\begin{gather}
0 \ = \ 2 \e^{K} \Big\{ 
\ol{W} D_t W - \frac{t - \ol{t}}{3} (\ol{D_t W})^2 
\Big\}
\, , \label{attract_eq}
\end{gather}
where the holomorphic central charge $W$ and its covariant derivative, and the symplectic invariant are given as
\bsubeq \label{D0D4_info}
\begin{gather}
W \ = \ q_0 + 3 p y^2
\, , \ls
D_t W \ = \ \frac{3}{2i y} \big( p y^2 - q_0 \big)
\, , \\
\VBH \ = \ 
|Z|^2 + G^{t \ol{t}} D_t Z \ol{D_t Z} \ = \ 
- \frac{1}{8 y^3} \Big\{ (q_0 + 3 p y^2)^2 + 3 (p y^2 - q_0)^2 \Big\}
\, .
\end{gather}
\esubeq
We are ready to look for a non-BPS solution which satisfies $D_t Z = \e^{K/2} D_t W \neq 0$ at the horizon.
Using (\ref{D0D4_info}), the attractor equation (\ref{attract_eq}) is given as
\begin{align}
0 \ &= \ 
\frac{3}{i y} \big( p y^2 - q_0 \big) \big( q_0 + p y^2 \big)
\, .
\end{align}
Since $D_t W \neq 0$ indicates $p y^2 \neq q_0$, we obtain 
\begin{align}
y^2 \ = \ - \frac{q_0}{p}
\end{align}
at the event horizon. 
Then the various values at the event horizon are given in the following forms:
\bsubeq \label{info_D0D4_nonBPS}
\begin{align}
Z \Big|_{r = \rH} \ &= \ 
- \frac{q_0}{\sqrt{2}} \Big( - \frac{p^3}{q_0^3} \Big)^{\frac{1}{4}}
\, , \ls
D_t Z \Big|_{r = \rH} \ = \ 
-3 i p \Big( - \frac{p}{q_0} \Big)^{\frac{1}{4}}
\, , \\
\VBH \Big|_{r = \rH} \ &= \ S_{\text{BH}}
\ = \ 
- 2 p^2 y \ = \ 
\sqrt{- 4 p^3 q_0}
\, .
\end{align}
\esubeq

\end{appendix}


}
\end{document}